\documentclass[aps,prb,reprint,footinbib,longbibliography]{revtex4-2}

\usepackage{graphicx}
\usepackage{subfigure}
\usepackage{bm}
\usepackage{amsmath}
\usepackage{amssymb}
\usepackage{url,hyperref}
\usepackage{siunitx}

\renewcommand{\vec}[1]{\bm{#1}}
\renewcommand{\d}[0]{\mathrm{d}} % roman d for derivatives and integrals
\newcommand{\e}{\mathrm{e}} % upright e for the natural number
\renewcommand{\Re}{\operatorname{Re}} % use roman instead of default script
\renewcommand{\Im}{\operatorname{Im}} % use roman instead of default script
\newcommand{\Tr}{\operatorname{Tr}} % trace

\begin{document}

\title{
    Clarification of Floquet-enhanced thermal emission through the nonequilibrium Green's function formalism
}

\author{Yuhua Ren} 
\email{yuhua.ren@u.nus.edu}
\affiliation{Department of Physics, 
National University of Singapore, 
Singapore 117551, 
Singapore}

\author{Hui Pan} 
\affiliation{Department of Physics, 
National University of Singapore, 
Singapore 117551, 
Singapore}

\author{Jian-Sheng Wang}
\email{phywjs@nus.edu.sg}
\affiliation{Department of Physics, 
National University of Singapore, 
Singapore 117551, 
Singapore}

%\date{10 June 2024}
\date{\today}
\bigskip

\begin{abstract}
Floquet engineering offers a powerful route to enhance emission in time-modulated media. 
Here, we investigate the influence of time-modulated permittivity in silicon carbide on its intensity spectrum. 
We consider both the nonequilibrium Green’s function approach and the macroscopic quantum electrodynamics approach, and establish their formal compatibility by deriving the Lippmann-Schwinger equation in both cases.
To analyze spectral features, we propose several methods for decomposing the electric field into positive- and negative-frequency components, along with the criteria required for physical consistency. 
Our analytical and numerical results show that, when defined appropriately, the intensity spectrum avoids divergence, though the resulting enhancement remains modest. 
These findings provide a unified theoretical foundation for modeling time-dependent media, and reinforce the utility of Floquet engineering as a versatile platform for tailoring emission dynamics.
\end{abstract}

\maketitle

\section{Introduction}

% Topic introduction (Heat transfer)
Radiative heat transfer has garnered significant attention as it is the fundamental mode of energy exchange in many natural and engineered systems. 
Control of thermal emission is crucial in numerous cutting-edge applications, including the thermal management of densely packed circuits, the suppression of infrared signatures for stealth technologies, and the thermal regulation of spacecraft in extreme environments.
Driven by these demands, recent research has focused on tailoring surface properties at micro- and nanoscales, enabling precise control over emissivity, polarization, and angular distribution of thermal radiation~\cite{greffet_incandescent_2024}. 
% Topic introduction (Floquet engineering)
Although spatial structuring has enabled remarkable control over thermal emission, emerging paradigms in temporal metamaterials suggest that modulating material properties in time can unlock novel regimes of radiative behavior~\cite{galiffi_wood_2020, ramaccia_phase-induced_2020, picardi_dynamic_2023, vazquez-lozano_incandescent_2023, yu_time-modulated_2024}. 
The introduction of time as a tunable degree of freedom redefines classical constraints such as energy conservation and reciprocity~\cite{ghanekar_nonreciprocal_2023}, enabling unprecedented control over light–matter interactions.
This shift not only challenges our existing understanding of radiative processes, but also highlights the pressing need for a theoretical framework capable of guiding future discoveries. 

% FE and MQED framework
At its core, the study of radiative heat transfer is a study of electromagnetic fields, governed by Maxwell's equations and shaped by material response, geometry, and boundary conditions.
A seminal contribution to the theoretical modeling of thermal radiation is Rytov's semiclassical theory of fluctuational electrodynamics (FE)~\cite{levin_theory_1967}. 
Rytov's theory treats the electromagnetic (EM) field classically, while incorporating quantum statistics through the fluctuation--dissipation theorem~\cite{callen_irreversibility_1951,kubo_fluctuation-dissipation_1966} to model stochastic thermal sources in matter.
Despite the success in explaining several key concepts such as near-field enhancements and the Casimir force~\cite{jones_thermal_2013}, FE is formulated for equilibrium situations and does not capture the full quantum aspect of EM fields~\cite{liberal_can_2025}.
Macroscopic quantum electrodynamics (MQED) is an extension of FE that quantizes the EM field in the presence of absorbing media~\cite{hopfield_theory_1958,scheel_macroscopic_2009}. 
Specifically, in the Huttner-Barnett model~\cite{huttner_dispersion_1992, huttner_quantization_1992, horsley_tutorial_2022}, the material is represented through a collection of harmonic oscillators coupled to the EM field, and diagonalizing the Hamiltonian leads to quantized polaritonic modes. 
The modern MQED framework successfully reproduces key results from FE and continues to propel progress in nanophotonics~\cite{feist_macroscopic_2021}, polariton chemistry~\cite{hsu_chemistry_2025}, and--most relevant to our work--radiative heat transfer~\cite{vazquez-lozano_incandescent_2023}.

% NEGF framework
% Overview of our results
In this paper, we apply the nonequilibrium Green's function (NEGF) method to analyze the electric field intensity spectrum of time-modulated silicon carbide (SiC). 
The NEGF method is a quantum many-body and nonequilibrium extension of FE, utilizing the Keldysh technique~\cite{henneberger_exact_2009, wang_transport_2023}. 
We demonstrate that several notable results previously reported in the literature~\cite{vazquez-lozano_incandescent_2023} can be recovered within the NEGF formalism through a comparatively streamlined derivation.
By employing second-order perturbation theory, analytical and numerical results for the Floquet-enhanced power spectrum were obtained. 
We confirmed the presence of Floquet-shifted peaks, but note that the magnitude may be significantly weaker than previously reported~\cite{vazquez-lozano_incandescent_2023}. 
We explain that the Bose-Einstein factor suppresses the tail of the spectrum, and how it is unnecessary to introduce dispersion or a high-frequency cutoff to address a known divergence issue~\cite{vertiz-conde_dispersion_2025}. 
We also take the opportunity to clarify the various definitions of the positive-frequency part of an operator, which might be unintuitive in the context of periodic modulation, where a positive frequency might be Floquet-shifted into a negative one and vice versa. 
Our work thus elucidate the subtle aspects of quantum statistics in Floquet systems and provide a comprehensive framework for analyzing time-modulated radiative phenomena.

The paper will be organized as follows. 
In section \ref{sec:negf}, we outline the NEGF method to model light-matter interactions, culminating in the Lippmann-Schwinger equation for perturbation theory. 
In section \ref{sec:mqed}, we extend the MQED Lagrangian to account for the time-varying medium, and arrived at the same Lippmann-Schwinger euqation as a consistency check. 
In section \ref{sec:discussion}, we obtain analytical and numerical expressions for the electric field intensity, and discuss the discrepencies with existing literature~\cite{vazquez-lozano_incandescent_2023}.  
\section{Nonequilibrium Green's function approach} \label{sec:negf}

\subsection{Photon Green's function definitions}

The photon NEGF offers a powerful framework for understanding radiative energy transfer, allowing a fully quantum analysis of photon-mediated interactions at the level of GFs and self-energies.
The photon NEGF is defined to be the two-point correlation function of the EM vector potential, but with the time argument defined on the Keldysh contour~\cite{keldysh_diagram_2023, stefanucci_nonequilibrium_2013} as
\begin{equation}
    D(\vec{r}, \tau, \vec{r}', \tau')_{\mu\nu} = \frac{1}{i \hbar} \langle
    \mathcal{T} A(\vec{r}, \tau)_{\mu} A(\vec{r}', \tau')_{\nu}
    \rangle \text{,}
\end{equation}
where $\mathcal{T}$ denotes contour-time ordering and $\langle \ldots \rangle$ represents the quantum expectation value.
For notational clarity, contour times are denoted by the Greek variable $\tau$, whereas physical times are denoted by the Latin letter $t$. 
Directional indices may be indicated in subscripts by other Greek symbols such as $\mu$ or $\nu$.
Here, $\vec{A}(\vec{r}, t)$ is to be treated as a quantum mechanical operator in the Heisenberg picture, unlike the case in FE. 
We adopt the temporal gauge, so that the electric field, $\vec{E} = - \partial_t \vec{A}$, and the magnetic field, $\vec{B} = \nabla \times \vec{A}$, can be expressed in terms of the vector potential $\vec{A}$ alone.
The different possibilities of the contour branches give us four definitions of GFs in physical time, of which not all are independent. 
The two important variants that describe the photon statistics are
\begin{subequations}
\begin{align}
    D^<(\vec{r}, t, \vec{r}', t')_{\mu\nu} &= \frac{1}{i \hbar} \langle
    A(\vec{r}', t')_{\nu} A(\vec{r}, t)_{\mu} 
    \rangle \text{,} \label{eq:d_less} \\
    D^>(\vec{r}, t, \vec{r}', t')_{\mu\nu} &= \frac{1}{i \hbar} \langle
    A(\vec{r}, t)_{\mu} A(\vec{r}', t')_{\nu} 
    \rangle \text{.} \label{eq:d_greater}
\end{align}
\end{subequations}
The retarded GF, which encodes the causal response, is defined as $D^R(t,t') = \Theta(t - t') [D^>(t,t') - D^<(t,t')]$~\cite{henneberger_exact_2009}, where position arguments are suppressed and $\Theta$ being the Heaviside step function. 
Subsequently, we will show that $D^R$ can be obtained by solving the Dyson equation~\cite{abrikosov_methods_2012} from a linear response perspective.

\subsection{Solving for the Green's functions}

In the FE community~\cite{dorofeyev_fluctuating_2011, joulain_surface_2005}, the response function relating the stochastic source currents and the EM field is called the dyadic Green's function. 
In the equilibrium case, our $D^R$ coincides with the electrical dyadic GF, modulo prefactors due to definitional conventions. 
According to Maxwell's equations, the relation between the vector potential and the total current is given by
\begin{equation}
     \vec{J} 
     =\frac{1}{\mu_0} (\nabla \times \nabla \times \vec{A} + \frac{1}{c^2} \partial_t^2 \vec{A}) 
     \equiv -v^{-1} \vec{A} 
     \text{.} \label{eq:v_inverse}
\end{equation} 
The second-order differential operator $v^{-1}$ admits a retarded GF, $v^R$~\cite{wang_transport_2023}. 
The minus sign convention ensures that the full photon GF, $D^R$, reduces to the vacuum version, $v^R$, when there is no medium present. 
In equilibrium, it is useful to take the Fourier transform in time, so that the time derivative becomes a multiplicative factor of $-i\omega$. 
The Fourier transform is thus defined with the convention $f(\omega) = \int \d t \, \exp(i \omega t) f(t)$. 
The current in Eq.~\eqref{eq:v_inverse} is the total current, and it can be decomposed as a free part $\vec{\xi}$, and an induced part arising from Ohm's law in a linear material, 
\begin{equation}
    \vec{J}(\vec{r}, \omega) = \vec{\xi}(\vec{r}, \omega) + \sigma(\vec{r}, \omega) i\omega \vec{A}(\vec{r}, \omega) 
    \text{.} \label{eq:ohms_law} 
\end{equation}
The link between conductivity and the dynamical behavior of the current operator is established through the Kubo formula.
In order to simplify the treatment while preserving the essential physics, a spatially local model will be used throughout this paper to avoid any added complications arising from spatial convolutions.
The material is also assumed to be nonmagnetic, with no bound current contribution due to magnetization. 

Suppose that the vector potential can be expressed as a linear response to the free current, i.e., 
\begin{equation}
    \vec{A}(\vec{r}, t) = - \int \d^3 \vec{r}_1 \int \d t_1 \, D^R(\vec{r}, t, \vec{r}_1, t_1) \vec{\xi}(\vec{r}_1, t_1) \text{.} \label{eq:dr_linear_response}
\end{equation}
When there is a low risk of misunderstanding, we may omit arguments and convolutions for brevity and write $\vec{A} = - D^R \vec{\xi}$ instead.
This also has the advantage that the equation has the same representation in both the time domain and the frequency domain. 
From Eqs.~\eqref{eq:v_inverse}--\eqref{eq:dr_linear_response}, $D^R$ must satisfy the Dyson equation~\cite{abrikosov_methods_2012}
\begin{align}
\begin{split}
    &v^{-1} D^R(\vec{r}, t, \vec{r}', t') = \delta^{(3)}(\vec{r}-\vec{r}') \delta(t-t') I \\
    &+ \int \d t_1 \int \d^3 \vec{r}_1 \, \Pi^R(\vec{r}, t, \vec{r}_1, t_1) D^R(\vec{r}, t_1, \vec{r}', t') \text{.} \label{eq:dyson_og}
\end{split}
\end{align}
The self-energy $\Pi^R = - i \omega \sigma$ is used in place of the conductivity, and it is also related to the equilibrium complex permittivity by $\Pi^R = - \epsilon_0 \omega^2 (\epsilon - 1)$~\cite{abrikosov_methods_2012}. 
In Eq.~\eqref{eq:ohms_law} and throughout, we assumed that the conductivity is local in space, i.e., $\Pi^R(\vec{r}, t, \vec{r}', t') = \Pi^R(\vec{r}, t, t') \delta^{(3)}(\vec{r}-\vec{r}')$. 
In the NEGF formalism, there are also other variants of the self-energy, such as $\Pi^<$, which relates to the current-current correlations.
Solving for the Dyson equation is a central step in the NEGF formalism, and it is usually carried out numerically except for the simplest geometries. 

In addition to $D^R$, there is also an advanced version, given by $D^A(t,t') = -\Theta(t' - t) [D^>(t,t') - D^<(t,t')]$. 
However, if $D^R$ is already solved, it does not require additional effort to obtain $D^A$ since 
\begin{equation}
    D^A(\vec{r}, t, \vec{r}', t') = D^R(\vec{r}', t', \vec{r}, t)^T 
    \text{.} \label{eq:da_from_dr}
\end{equation}
Thus, we have the necessary ingredients needed to calculate the lesser variant of the photon GF, $D^<$, by the Keldysh equation~\cite{kruger_trace_2012, aeberhard_photon_2014, wang_transport_2023}
\begin{align}
\begin{split}
    &D^<(\vec{r}, t, \vec{r}', t') = \int \d^3 \vec{r}_1 \int \d t_1 \int \d^3 \vec{r}_2 \int \d t_2 \\
    &\times D^R(\vec{r}, t, \vec{r}_1, t_1) \Pi^<(\vec{r}_1, t_1, \vec{r}_2, t_2) D^A(\vec{r}_2, t_2, \vec{r}', t') 
    \text{.} \label{eq:keldysh_eq}
\end{split}
\end{align}
Here, $\Pi^<$ is the lesser self-energy in the presence of the material, and in equilibrium, it is related to the retarded and advanced variants by the fluctuation--dissipation theorem, $\Pi^< = N (\Pi^R - \Pi^A)$, where $N(\omega) = [\exp(\hbar\omega/k_B T) - 1]^{-1}$ is the Bose-Einstein function at temperature $T$. 
The advanced self-energy is obtained by $\Pi^A = (\Pi^R)^\dagger$, analogous to the case for $D^A$ in Eq.~\eqref{eq:da_from_dr}.
The lesser self-energy also has the interpretation of 
\begin{equation}
    \Pi^<(\vec{r}, t, \vec{r}', t')_{\mu\nu} 
    = \frac{1}{i \hbar} \langle
    \xi(\vec{r}', t')_{\nu} \xi(\vec{r}, t)_{\mu} 
    \rangle 
    \text{,} \label{eq:pi_less}
\end{equation} 
from which the Keldysh equation follows by applying Eq.~\eqref{eq:dr_linear_response} to each copy of $\xi$ and arriving at Eq.~\eqref{eq:d_less}.

\subsection{Floquet drive}

The temporal modulation in permittivity necessarily corresponds to a $\Pi^R$ that is non-local in time. 
Consider the modification $\Pi^R \to \Pi^R_{\textrm{eq}} + \Pi_{\textrm{dr}}$, where it is assumed $\Pi_{\textrm{dr}}^\dagger = \Pi_{\textrm{dr}}$.
This ensures that $\Pi^A \to \Pi^A_{\textrm{eq}} + \Pi_{\textrm{dr}}$, and $\Pi^<$, or equivalently $\Im \epsilon$, remains unchanged from its equilibrium version. 
This important assumption implies that the fluctuation--dissipation theorem for the stochastic current $\vec{\xi}$ still holds under time modulation~\cite{yu_manipulating_2023}. 
The time modulation breaks time-translational symmetry, rendering the temporal Fourier transform inapplicable.
However, the infinite-dimensional Floquet representation enables the Floquet convolution theorem~\cite{tsuji_correlated_2008} when the perturbation is periodic. 
For a function with two arguments that satisfies the discrete time translation symmetry of period $2\pi/\Omega$, $G(t,t') = G(t+2\pi/\Omega, t'+2\pi/\Omega)$, its Floquet representation is defined as 
\begin{align}
    &G(t, t') \to G_{mn}(\omega) \notag \\
    &= \frac{\Omega}{2\pi} \int_0^{\frac{2\pi}{\Omega}} \d t \int_{-\infty}^\infty \d t' \e^{i (\omega_m t - \omega_n t')} G(t, t') \text{,} 
\end{align}
where $\omega_m = \omega + m \Omega$.
Floquet matrices will be denoted in bold; their distinction from vectors sharing the same notation is typically evident from context. 
The Floquet convolution theorem is then the Floquet representation of time convolutions, namely,  
\begin{equation}
    \int \d t_1 \, A(t, t_1) B(t_1, t') \to \sum_{k} A_{mk}(\omega) B_{kn}(\omega) \text{.} 
\end{equation}
Since convolution is generally a noncommutative operation, the order of multiplying Floquet matrices should be treated with caution. 
The relation between the time-modulated component of the self-energy and the time-modulation of the susceptibility is given by~\cite{zhu_enhancing_2025}
\begin{subequations}
\begin{align}
    \Pi_{\textrm{dr}}(t, t') &= -\epsilon_0 \partial_t \partial_{t'} \chi_{\textrm{dr}}(t, t') 
    \text{,} \label{eq:driven_pi} \\
    \bm{\Pi}_{\textrm{dr}}(\omega) &= - \epsilon_0 \bm{\Omega}(\omega) \bm{\chi}_{\textrm{dr}}(\omega) \bm{\Omega}(\omega) 
    \text{,} \label{eq:driven_pi_floquet}
\end{align}
\end{subequations}
where $\bm{\chi}_{\textrm{dr}}(\omega)$ is the Floquet representation of the modulation $\chi(t,t')$.
The diagonal Floquet matrix $\bm{\Omega}$ is the Floquet counterpart to the time derivatives and has matrix elements $\bm{\Omega}_{mn} = \delta_{mn} \omega_n I$. 
Another way to write the two-time susceptibility is by performing a double Fourier transform to obtain its two-frequency representation~\cite{sloan_casimir_2021,yu_manipulating_2023}.

As a concrete application, we consider the same setup as in the work titled ``Incandescent temporal metamaterials''~\cite{vazquez-lozano_incandescent_2023}. 
A semi-infinite SiC slab occupies the region $z<0$, with a vacuum on the other side.
The permittivity of SiC, $\epsilon(t,t')$, is modeled as its equilibrium value, $\epsilon_{\textrm{eq}}(t-t')$, with an additional contribution arising from the drive, $\chi_{\textrm{dr}}(t,t')$.
The equilibrium permittivity is based on the Lorentz model, with the frequency dependence given by 
\begin{equation}
    \epsilon_{\textrm{eq}}(\omega) = \epsilon_{\infty} \frac{\omega_L^2 -\omega^2 - i\gamma \omega}{\omega_T^2 - \omega^2 - i \gamma \omega} \text{,} 
\end{equation}
with the values $\epsilon_\infty = 6.7$, $\omega_L/2\pi = \SI{29.1}{\tera\hertz}$, $\omega_T/2\pi = \SI{23.8}{\tera\hertz}$ , and $\gamma/2\pi =  \SI{0.14}{\tera\hertz}$. 
The driven part is sinusoidal and local in time, to reflect the periodical and active change in the refractive index,  
\begin{subequations}
\begin{align}
    \chi_{\textrm{dr}}(\vec{r}, t, \vec{r}', t') &= \chi_{\textrm{dr}}(t) \delta^{(3)}(\vec{r}-\vec{r}') \delta(t-t') \text{,} \\
    \chi_{\textrm{dr}}(t) &= 2 \Delta \chi \cos(\Omega t) 
    \text{.} \label{eq:sinusoidal_modulation}
\end{align}
\end{subequations}
The modulation parameters assumed in subsequent numerical calculations are $\Omega/2\pi = \SI{1.5}{\tera\hertz}$ and $\Delta \chi = 0.025$.
Anticipating later usage, the matrix elements for the corresponding $\bm{\Pi}_{\textrm{dr}}$ are worked out to be
\begin{equation}
    \Pi_{\textrm{dr},mn}(\omega) = -\epsilon_0 \omega_m \omega_n \Delta \chi \delta_{|m-n|,1} \text{.} \label{eq:pi_dr_floquet}
\end{equation}
As evident in the Floquet representation, the off-diagonal nature of $\Pi_{\textrm{dr}}$ is responsible for the Floquet shift of $\pm \Omega$. 
This particular choice of time modulation is in line with our earlier remark of not perturbing the imaginary part of $\Pi^R$.

\subsection{Lippmann-Schwinger equation}

In addition to NEGF being a powerful mathematical framework on its own, it also provides an iterative way of calculating quantities of interest.
Here, we demonstrate how the electric field in the driven case can be calculated from the equilibrium electric field by the Lippmann-Schwinger equation~\cite{lippmann_variational_1950,herz_green-kubo_2019}. 

To proceed, we make the distinction regarding quantities without time modulation by having the subscript ``eq''.
The undriven and driven versions of $D^R$ each satisfy their own Dyson equations, which are respectively written as 
\begin{subequations}
\begin{align}
    (v^{-1} - \Pi^R_{\textrm{eq}}) D^R_{\textrm{eq}} &= I \text{,} \\
    (v^{-1} - \Pi^R_{\textrm{eq}}) D^R &= I + \Pi_{\textrm{dr}} D^R \text{.} 
\end{align}
\end{subequations}
Upon eliminating $(v^{-1} - \Pi^R_{\textrm{eq}})$, a new Dyson equation involving $D^R$ and $D^R_{\textrm{eq}}$ is obtained, 
\begin{equation}
    D^R = D^R_{\textrm{eq}} + D^R_{\textrm{eq}} \Pi_{\textrm{dr}} D^R 
    \text{.} \label{eq:dyson_ls}
\end{equation}

As an analog of Eq.~\eqref{eq:dr_linear_response}, we can define an equilibrium linear response relation by $\vec{A}_{\textrm{eq}} = - D^R_{\textrm{eq}} \vec{\xi}$.
By acting Eq.~\eqref{eq:dyson_ls} on $-\vec{\xi}$, $\vec{A}$ is seen to be related to $\vec{A}_{\textrm{eq}}$ by the Lippmann-Schwinger equation~\cite{henneberger_exact_2009}, 
\begin{equation}
    \vec{A} = \vec{A}_{\textrm{eq}} + D^R_{\textrm{eq}} \Pi_{\textrm{dr}} \vec{A} 
    \text{.} \label{eq:ls_a}
\end{equation}
Using Eq.~\eqref{eq:driven_pi} and performing integration by parts, the Lippmann-Schwinger equation for the electric field is obtained~\cite{sloan_casimir_2021},
\begin{equation}
\begin{split}
    &\vec{E}(\vec{r}, t) 
    = \vec{E}_{\textrm{eq}}(\vec{r}, t) + \int \d^3 \vec{r}_1 \int_{-\infty}^{\infty} \d t_1 \, \\
    & \times D^R_{\textrm{eq}}(\vec{r}, \vec{r}_1, t-t_1)[\partial_{t_1}^2  \epsilon_0 \chi_{\textrm{dr}}(\vec{r}_1, t_1)] \vec{E}(\vec{r}_1, t_1) 
    \text{.} \label{eq:ls_e}
\end{split} 
\end{equation}
Due to time translational invariance, the equilibrium GF will be a function of the difference in the time arguments, and can thus be Fourier transformed.
In the Floquet representation, they are diagonal matrices, which enables them to commute with other diagonal matrices. 
As we shall see later, the same Lippmann-Schwinger equation obtained using NEGF can also be derived using MQED, thus further reinforcing our confidence in the validity of both theories.

\subsection{Perturbation theory}

Although a closed-form solution may not be tractable for the Floquet-driven system, meaningful results can still be obtained by considering a perturbative approach.
Here, we assume that the modulation is sufficiently weak, and truncate the Born series given by Eq.~\eqref{eq:dyson_ls} to second order in $\Pi_{\textrm{dr}}$, 
\begin{equation}
    D^R \approx [I 
    + D^R_{\textrm{eq}} \Pi_{\textrm{dr}} 
    + (D^R_{\textrm{eq}} \Pi_{\textrm{dr}})^2] D^R_{\textrm{eq}} \text{.} \label{eq:dyson_truncated}
\end{equation}
The same can be done for $D^A$. 
The Keldysh equation, Eq.~\eqref{eq:keldysh_eq}, is then used to obtain the lesser GF, which serves as the field correlations.
To second order in $\Pi_{\textrm{dr}}$, we have 
\begin{align}
\begin{split}
    &D^< \approx D^<_{\textrm{eq}} 
    + D^R_{\textrm{eq}} \Pi_{\textrm{dr}} D^<_{\textrm{eq}} 
    + D^<_{\textrm{eq}} \Pi_{\textrm{dr}} D^A_{\textrm{eq}} \\
    &+ (D^R_{\textrm{eq}} \Pi_{\textrm{dr}})^2 D^<_{\textrm{eq}} 
    + D^R_{\textrm{eq}} \Pi_{\textrm{dr}} D^<_{\textrm{eq}} \Pi_{\textrm{dr}} D^A_{\textrm{eq}} \\
    &+ D^<_{\textrm{eq}} (\Pi_{\textrm{dr}} D^A_{\textrm{eq}})^2 \text{.}
\end{split} \label{eq:keldysh_truncated}
\end{align}
The fluctuation--dissipation theorem is encapsulated in $D^<_{\textrm{eq}}$, since
\begin{equation}
    D^<_{\textrm{eq}}(\omega) = -2i \omega^2 \epsilon_0 N(\omega) D^R_{\textrm{eq}}(\omega) \Im[\epsilon_{\textrm{eq}}(\omega)] D^A_{\textrm{eq}}(\omega) \text{.}
\end{equation}

Observe that all terms in Eq.~\eqref{eq:keldysh_truncated} are proportional to $D^<_{\textrm{eq}}$, which implies that $D^<$ is proportional to $N$, the Bose-Einstein function. 
This conclusion remains true even when considering higher-order terms in the perturbation series. 
The structure of Eq.~\eqref{eq:keldysh_truncated} can also be interpreted as an application of Langreth's rules~\cite{stefanucci_nonequilibrium_2013}. 
In Eq.~\eqref{eq:keldysh_truncated}, all the Floquet matrices are diagonal except $\bm{\Pi}_{\textrm{dr}}$, which is tridiagonal but zero along the main diagonal. 
It becomes clear that these off-diagonal elements are responsible for the $\pm \Omega$ Floquet shifts, and we would expect $\pm 2\Omega, \pm 3 \Omega, \ldots$ shifts if more terms in the perturbation series are kept. 

A plethora of interesting quantities involving a product of the fields can be calculated from $D^<$.
Of immediate interest is the energy density of the electric field, which can be obtained by evaluating $D^<$ at coincident points, 
\begin{equation}
    \langle \vec{E} \cdot \vec{E} \rangle = i\hbar \sum_{\mu} \bm{\Omega} \bm{D}^<_{\mu\mu} \bm{\Omega} \text{.} \label{eq:ee_from_dless}
\end{equation} 
This expression will be further elaborated and subsequently employed in numerical calculations.
In the next section, we turn to a quantitative analysis of $\langle E E \rangle$ using the MQED approach to check for any inconsistencies between the two theories. 

\section{Macroscopic quantum electrodynamics approach} \label{sec:mqed}

\subsection{MQED Hamiltonian}

The MQED approach is a powerful extension to FE in the presence of lossy materials. 
The starting point for MQED is the Lagrangian, which collectively describes the field, matter, and their interaction.
In the Huttner-Barnett model, the unperturbed Hamiltonian is the sum of the energies of the electromagnetic field and a collection of harmonic oscillators which serve as the bath. 
The bath is essential to introduce dissipation in the model of a lossy dielectric. 
In addition to the Lagrangian previously studied~\cite{horsley_tutorial_2022}, we introduce a new term modulating the electric field density to produce the desired time modulation of the permittivity.
The total Lagrangian is thus given as
\begin{equation}
    \mathcal{L} = \mathcal{L}_{\textrm{f}} + \mathcal{L}_{\textrm{r}} + \mathcal{L}_{\textrm{i}} + \mathcal{L}_{\textrm{dr}} \text{,}
\end{equation}
where the terms correspond to the fields, reservoir, interaction, and drive, respectively.
They are defined as follows:
\begin{subequations}
\begin{align}
    \mathcal{L}_{\textrm{f}} &= \frac{1}{2} \left( \epsilon_0 |\vec{E}|^2 - \frac{1}{\mu_0} |\vec{B}|^2 \right) \text{,} \label{eq:mqed_lagrangian_f} \\
    \mathcal{L}_{\textrm{r}} &= \frac{1}{2} \int_0^\infty \d\omega\, \biggl[ \left(\partial_t \vec{X}^\omega \right)^2 - \omega^2 (\vec{X}^\omega)^2 \biggr] \text{,} \\
    \mathcal{L}_{\textrm{i}} &= \vec{E} \cdot \int_0^\infty \d\omega\, \alpha(\omega) \vec{X}^\omega \text{,} \label{eq:mqed_lagrangian_i} \\
    \mathcal{L}_{\textrm{dr}} &= -\frac{1}{2} \vec{E} \cdot \vec{P}_{\textrm{dr}} \text{,}
\end{align}
\end{subequations}
where $\vec{P}_{\textrm{dr}}(\vec{r}, t) = \epsilon_0 \chi_{\textrm{dr}}(\vec{r}, t) \vec{E}(\vec{r}, t)$ is the additional polarization induced by the time-varying susceptibility.
The factor $\alpha$ represents the material's polarizability~\cite{horsley_tutorial_2022} and is related to the lossy part of the dielectric permittivity by
\begin{equation}
    \alpha(\vec{r}, \omega) = \sqrt{2 \omega \Im[\epsilon_0 \epsilon_{\textrm{eq}}(\vec{r}, \omega)] / \pi} \text{.}
\end{equation}

Although the Lagrangian is written in terms of the electric field and the magnetic field, the field variables should be understood to be only $\vec{X}^\omega$ and $\vec{A}$. 
Solving the Euler-Lagrange equations then yields the following equations of motion, 
\begin{subequations}
\begin{align}
    - v^{-1} \vec{A} &= \partial_t \left( \vec{P}_{\textrm{dr}}
    + \int_0^\infty \d\omega\, \alpha(\omega) \vec{X}^\omega \right) \text{,} \label{eq:eom_a} \\
    \alpha(\omega) \vec{E} &= \left(\omega^2 + \partial_t^2 \right)  \vec{X}^\omega 
    \text{.} \label{eq:eom_x}
\end{align}
\end{subequations} 
The microscopic variable $\vec{X}^\omega$ can be eliminated by solving the simple harmonic oscillator equation, Eq.~\eqref{eq:eom_x}, leading to
\begin{equation}
    \vec{X}^\omega(\vec{r}, t) = \int_{-\infty}^{\infty} \d t_1 \, g^\omega(t-t_1) \alpha(\vec{r}, \omega) \vec{E}(\vec{r}, t_1) + \vec{X}^\omega_0(\vec{r}, t) \text{,}
\end{equation}
where $g^\omega(t) = \omega^{-1} \Theta(t) \sin(\omega t)$ is the GF for the differential operator $(\omega^2 + \partial_t^2)$, and $\vec{X}_0^\omega$ is the homogeneous solution. 
The left-hand side of Eq.~\eqref{eq:eom_a} is the total current according to Maxwell's equations, and on the right-hand side, the free current is contributed by the homogeneous solution of $\vec{X}^\omega$, 
\begin{equation}
    \vec{\xi}(\vec{r}, t) = \partial_t \int_0^\infty \d \omega \, \alpha(\vec{r}, \omega) \vec{X}^\omega_0(\vec{r}, t) . \label{eq:mqed_free_current}
\end{equation}

The homogeneous solution is a linear combination of complex exponentials, so $\vec{X}_0^\omega$ can be further expressed in terms of the polaritonic operators $\vec{a}_{0}^\omega$ and $(\vec{a}_{0}^\omega)^\dagger$, 
\begin{equation}
    \vec{X}^\omega_0(\vec{r}, t) = \sqrt{\frac{\hbar}{2\omega}} \biggl[\vec{a}_{0}^{\omega}(\vec{r}) \e^{-i\omega t} + \vec{a}_{0}^{\omega}(\vec{r})^\dagger \e^{i\omega t}\biggr] 
    \text{,} \label{eq:qho_position}
\end{equation}
which satisfy 
\begin{subequations}
\begin{align}
    [a_{0}^{\omega}(\vec{r})_\mu, a_{0}^{\omega'}(\vec{r}')_{\mu'}^\dagger] &= \delta_{\mu\mu'} \delta^{(3)}(\vec{r}-\vec{r}') \delta(\omega - \omega') 
    \text{,} \label{eq:ccr} \\
    \langle a_{0}^{\omega}(\vec{r})_\mu^\dagger a_{0}^{\omega'}(\vec{r}')_{\mu'} \rangle &= \delta_{\mu\mu'} \delta^{(3)}(\vec{r}-\vec{r}') \delta(\omega - \omega') N(\omega) \text{.} \label{eq:ccr_statistics}
\end{align}
\end{subequations}
Eq.~\eqref{eq:qho_position} is the standard transformation used to express the position operator for the quantum harmonic oscillator in terms of creation and annihilation operators. 
To be further convinced of the validity of the equal-time commutation relations, the free current-current correlation can be worked out to give Eq.~\eqref{eq:pi_less} exactly, which in equilibrium is nothing but the fluctuation--dissipation theorem~\cite{song_near-field_2015}. 

To get to the Hamiltonian description, the canonical momentum variables are obtained by differentiating the Lagrangian, and they are found to be: 
\begin{subequations}
\begin{align}
\begin{split}
    \vec{\Pi}_{A} &= - \epsilon_0 \vec{E} + \vec{P}_{\textrm{dr}} 
    - \int_0^\infty \d\omega\, \alpha(\omega) \vec{X}^\omega \text{,} 
\end{split} \\
    \vec{\Pi}_{X} &= \partial_t \vec{X}^\omega \text{.}
\end{align}
\end{subequations}
The Legendre transform is then used to get the Hamiltonian from the Lagrangian, given by
\begin{align}
    \mathcal{H} &= \vec{\Pi}_A \cdot \partial_t \vec{A} + \int_0^\infty \d \omega \, \vec{\Pi}_X \cdot \partial_t \vec{X}^\omega - \mathcal{L} \notag \\
    &= -\frac{1}{2} \vec{E} \cdot \vec{P}_{\textrm{dr}} + \frac{1}{2} \left( \epsilon_0 |\vec{E}|^2 + \frac{1}{\mu_0} |\vec{B}|^2 \right) \notag \\
    &\quad + \frac{1}{2} \int_0^\infty \d\omega\, \biggl[ \left(\partial_t \vec{X}^\omega \right)^2 + \omega^2 (\vec{X}^\omega)^2 \biggr] \text{.}
\end{align}

The resulting Hamiltonian can be expressed as $\mathcal{H} = \mathcal{H}_0 + \mathcal{H}_T$, where $\mathcal{H}_0$ is the unperturbed term corresponding to the energy in the oscillators and fields, and $\mathcal{H}_T  = -\tfrac{1}{2} \vec{E} \cdot \vec{P}_{\textrm{dr}}$ is the interaction term due to time modulation.
In contrast to previous works~\cite{sloan_casimir_2021, vazquez-lozano_incandescent_2023}, the discrepancy in the factor of half can be understood as the energy built up from the accumulation of an induced dipole moment, in contrast to a permanent dipole.
Nevertheless, such global factors or phase differences can be absorbed into $\Delta \chi$ in Eq.~\eqref{eq:sinusoidal_modulation} for ease of cross-referencing. 
Specifically, $\Delta \chi$ here is defined to be half the modulation amplitude. 
We note parenthetically that a recent study has explored the modeling of a time-varying medium with MQED by driving the reservoir instead~\cite{horsley_macroscopic_2025}.

After quantization~\cite{fradkin_quantum_2021} and performing a diagonalization procedure~\cite{fano_atomic_1956,suttorp_fano_2004,scheel_macroscopic_2009,philbin_canonical_2010}, the unperturbed Hamiltonian can be written in terms of polaritonic operators, $\vec{a}$ and $\vec{a}^\dagger$, 
\begin{equation}
\begin{split}
    \mathcal{H}_0(t) &= \int_0^{\infty} \d \omega \, \frac{\hbar\omega}{2} \sum_{\mu} \biggl[ \\
    &\quad a^\omega(\vec{r}, t)_\mu^\dagger a^\omega(\vec{r}, t)_\mu + a^\omega(\vec{r}, t)_\mu a^\omega(\vec{r}, t)_\mu^\dagger \biggr] \text{.}
\end{split}
\end{equation}
As a reminder, the operator $a$ carries time dependence in the Heisenberg picture. 
In equilibrium, the operators are free-evolving,  
\begin{equation}
    \vec{a}^\omega(\vec{r}, t) = \vec{a}_0^\omega(\vec{r}) \e^{-i \omega t} 
    \text{.} \label{eq:time_evo_a_eq}
\end{equation}
In the driven case, the time dependence is more complicated, but an implicit relation can be obtained by turning to the Heisenberg equation of motion.

\subsection{Expression for electric field}

Here, we derive an expression for the electric field operator in terms of the polaritonic operators. 
This effort is motivated, in part, by the necessity of computing electric field commutation relations that arise later through the Heisenberg equation.
Using Eqs.~\eqref{eq:mqed_free_current} and \eqref{eq:qho_position}, the free current can be first expressed in terms of the polaritonic operators. 
According to our NEGF theory, the vector potential can be obtained from the free current via Eq.~\eqref{eq:dr_linear_response}, which results in 
\begin{equation}
\begin{split}
    &\vec{A}(\vec{r}, t) = -\int \d^3 \vec{r}_1 \int \d t_1 \, D^R_{\textrm{eq}}(\vec{r}, \vec{r}_1, t - t_1) \int_0^\infty \d \omega \\
    &\times \alpha(\vec{r}_1, \omega) i \sqrt{\frac{\hbar \omega}{2}} \biggl[-\vec{a}_0^{\omega}(\vec{r}_1) \e^{-i \omega t_1} + \vec{a}_0^{\omega}(\vec{r}_1)^\dagger \e^{i\omega t_1} \biggr] \text{.}
\end{split}
\end{equation} 
Taking the time derivative and applying integration by parts gives the expression for the electric field, 
\begin{equation}
\begin{split}
    &\vec{E}(\vec{r}, t) = - \int \d^3 \vec{r}_1 \int \d t_1 \, D^R_{\textrm{eq}}(\vec{r}, \vec{r}_1, t - t_1) \int_0^\infty \d \omega \\
    &\times \alpha(\vec{r}_1, \omega) \sqrt{\frac{\hbar \omega}{2}} \omega \biggl[\vec{a}_0^{\omega}(\vec{r}_1) \e^{-i \omega t_1} + \vec{a}_0^{\omega}(\vec{r}_1)^\dagger \e^{i\omega t_1} \biggr] 
    \text{.}
\end{split}
\end{equation}
The difficulty arises from the time convolution, which requires knowing the polaritonic operators for all times in the past and into the future. 
We instead take the Fourier transform in the variable $t_1$ to obtain
\begin{equation}
\begin{split}
    \vec{E}(\vec{r}, t) = - \int \d^3 \vec{r}_1 \int_0^\infty \d \omega \, \alpha(\vec{r}_1, \omega) \sqrt{\frac{\hbar \omega}{2}} \omega \\
    \times \biggl[D^R_{\textrm{eq}}(\vec{r}, \vec{r}_1, \omega) \vec{a}_0^{\omega}(\vec{r}_1) \e^{-i \omega t} \\
    + D^R_{\textrm{eq}}(\vec{r}, \vec{r}_1, -\omega) \vec{a}_0^{\omega}(\vec{r}_1)^\dagger \e^{i\omega t} \biggr] 
    \text{.} \label{eq:pre_mqed_e}
\end{split}
\end{equation}
Here, we identify the time-dependent polaritonic operator using Eq.~\eqref{eq:time_evo_a_eq}.
We can also clearly see that Eq.~\eqref{eq:pre_mqed_e} separates into positive and negative $\omega$ terms. 
We are thus motivated to define the electric field operators as a linear combination of the polaritonic operators,
\begin{subequations}
\begin{align}
\begin{split}
    \vec{E}(\vec{r}, t)^{(+)} 
    &= - \int \d^3 \vec{r}_1 \int_0^\infty \d \omega \, \alpha(\vec{r}_1, \omega) \sqrt{\frac{\hbar \omega}{2}} \omega \\
    &\quad \times D^R_{\textrm{eq}}(\vec{r}, \vec{r}_1, \omega) \vec{a}^{\omega}(\vec{r}_1, t) 
    \text{,} \label{eq:mqed_e_plus}
\end{split} \\
\begin{split}
    \vec{E}(\vec{r}, t)^{(-)} 
    &= - \int \d^3 \vec{r}_1 \int_0^\infty \d \omega \, \alpha(\vec{r}_1, \omega) \sqrt{\frac{\hbar \omega}{2}} \omega \\
    &\quad \times D^R_{\textrm{eq}}(\vec{r}, \vec{r}_1, -\omega) \vec{a}^{\omega}(\vec{r}_1, t)^\dagger 
    \text{.} \label{eq:mqed_e_minus} 
\end{split}
\end{align}
\end{subequations}
This is seen to be identical to Eq.~(4) in Ref.~\cite{vazquez-lozano_incandescent_2023}, by substituting $D^R_{\textrm{eq}} = - \mu_0 G$ and $\vec{a} = i \hat{\vec{f}}$.  
The total electric field is then the sum of the positive-frequency component and the negative-frequency component, $\vec{E} = \vec{E}^{(+)} + \vec{E}^{(-)}$. 
From Eq.~\eqref{eq:mqed_e_plus} (Eq.~\eqref{eq:mqed_e_minus}), it may be tempting to conclude that $E^{(+)}$ ($E^{(-)}$) only contains the annihilation (creation) operator, but that is incorrect since $\vec{a}^\omega(\vec{r}, t)$ in the Heisenberg picture consists of both $\vec{a}^\omega_0$ and $(\vec{a}^\omega_0)^\dagger$ in general.
In contrast to the static case, it will be seen that the time-dependent driving results in the mixing of both operators. 

\subsection{Heisenberg equation and time evolution}

Unlike the diagonal case, where the operator evolution simply means acquiring a phase, the situation here is complicated by the additional interaction term. 
Nevertheless, the Heisenberg equation can be used to determine the time evolution of the full polaritonic operators, 
\begin{equation}
    i\hbar \partial_t \vec{a}^\omega(\vec{r}, t) 
    = \biggl[\vec{a}^\omega(\vec{r}, t), H_0(t) + H_T(t)\biggr] 
    \text{.} \label{eq:heisenberg_a}
\end{equation}

The first term in the commutator is rather straightforward, giving 
\begin{equation}
    \biggl[a^\omega(\vec{r}, t)_\mu, H_0(t)\biggr] = \hbar \omega a^\omega(\vec{r}, t)_\mu 
    \text{.}
\end{equation}
The second term involves more effort, as we need to simplify 
\begin{equation}
\begin{split}
    \biggl[a^\omega(\vec{r}, t)_\mu, H_T(t)\biggr] &= -\int \d^3 \vec{r}_1 \, \frac{\epsilon_0}{2} \chi_{\textrm{dr}}(t) \\
    &\qquad \times \biggl[\vec{a}^\omega(\vec{r}, t), E(\vec{r}_1, t)^2\biggr] 
    \text{.}
\end{split}
\end{equation}
To work out the inner commutator, Eq.~\eqref{eq:mqed_e_plus} and Eq.~\eqref{eq:mqed_e_minus} are invoked to yield
\begin{equation}
\begin{split}
    \biggl[a^\omega(\vec{r}, t)_\mu, E(\vec{r}', t)^2 \biggr] \\
    = -2 \sum_{\nu} E(\vec{r}', t)_{\nu} D^A_{\textrm{eq}}(\vec{r}, \vec{r}', \omega)_{\mu\nu} \alpha(\vec{r}, \omega) \sqrt{\frac{\hbar\omega}{2}} \omega \text{.}
\end{split}
\end{equation}
The switch from $D^R$ to $D^A$ is done using the Fourier transform of Eq.~\eqref{eq:da_from_dr}, $D^A_{\textrm{eq}}(\vec{r}, \vec{r}', \omega) = D^R_{\textrm{eq}}(\vec{r}', \vec{r}, -\omega)^T$. 
Eq.~\eqref{eq:heisenberg_a} can then be reorganized so that all the $\vec{a}^\omega(\vec{r}, t)$ terms are on one side, 
\begin{equation}
\begin{split}
    (i \partial_t - \omega) a^\omega(\vec{r}, t)_\mu 
    = \int \d^3 \vec{r}_1 \, \epsilon_0 \chi_{\textrm{dr}}(t) \\
    \times \sum_{\nu} E(\vec{r}_1, t)_{\nu} D^A_{\textrm{eq}}(\vec{r}, \vec{r}_1, \omega)_{\mu\nu} \alpha(\vec{r}, \omega) \sqrt{\frac{\omega}{2 \hbar}} \omega \text{.}
\end{split}
\end{equation}
This is a first-order differential equation like the Schr\"{o}dinger equation, and the solution can be obtained by knowing the GF, $(i\partial_t - \omega) [-i \Theta(t) \e^{-i \omega t}] = \delta(t)$.
The full solution to Eq.~\eqref{eq:heisenberg_a} is thus
\begin{equation}
\begin{split}
    &a^\omega(\vec{r}, t)_{\mu} 
    = a^\omega_0(\vec{r})_{\mu} \e^{-i\omega t} \\
    &- i \alpha(\vec{r},\omega) \sqrt{\frac{\omega}{2\hbar}} \omega \int_{-\infty}^t \d t_1 \, \e^{-i \omega (t-t_1)} \int \d^3 \vec{r}_1 \\
    &\times \sum_{\nu} D^A_{\textrm{eq}}(\vec{r}, \vec{r}_1, \omega)_{\mu\nu} \epsilon_0 \chi_{\textrm{dr}}(t_1) E_{\nu}(\vec{r}_1, t_1)
    \text{.} \label{eq:a_soln}
\end{split}
\end{equation}
The first term is the homogeneous solution and can be interpreted as the solution to the undriven case by Eq.~\eqref{eq:time_evo_a_eq} when $\chi_{\textrm{dr}}$ is set to $0$.

\subsection{Lippmann-Schwinger equation}

With Eqs.~\eqref{eq:mqed_e_plus} and \eqref{eq:mqed_e_minus}, we can construct the equilibrium electric field, $\vec{E}_{\textrm{eq}}$ (associated with the homogeneous solution), and the actual electric field, $\vec{E}$ (associated with the full solution). 
In the intermediate steps, we use the completeness relation~\cite{vazquez-lozano_incandescent_2023}, $D^R_{\textrm{eq}} (\Pi^R_{\textrm{eq}} - \Pi^A_{\textrm{eq}}) D^A_{\textrm{eq}} = D^R_{\textrm{eq}} - D^A_{\textrm{eq}}$. 
In a similar fashion to Eq.~\eqref{eq:a_soln}, the actual electric field differs from the equilibrium electric field by
\begin{subequations}
\begin{align}
\begin{split}
    &\vec{E}^{(+)}(\vec{r}, t) 
    = \vec{E}_{\textrm{eq}}^{(+)}(\vec{r}, t) - \int \d^3 \vec{r}_1 \int_{-\infty}^{t} \d t_1 \, \\
    &\times K(\vec{r}, \vec{r}_1, t-t_1) \epsilon_0 \chi_{\textrm{dr}}(t_1) \vec{E}(\vec{r}_1, t_1) 
    \text{,} \label{eq:pre_ls_plus}
\end{split} \\
\begin{split}
    &\vec{E}^{(-)}(\vec{r}, t) 
    = \vec{E}_{\textrm{eq}}^{(-)}(\vec{r}, t) - \int \d^3 \vec{r}_1 \int_{-\infty}^{t} \d t_1 \, \\
    &\times K(\vec{r}, \vec{r}_1, t-t_1)^* \epsilon_0 \chi_{\textrm{dr}}(t_1) \vec{E}(\vec{r}_1, t_1) 
    \text{.} \label{eq:pre_ls_minus}
\end{split}
\end{align}
\end{subequations}
The kernel $K$ is given by
\begin{equation}
\begin{split}
    K(\vec{r}, \vec{r}', t) &= \int_0^\infty \frac{\d \omega}{2\pi} \, \e^{-i \omega t} \omega^2 \\
    &\quad \times \biggl[D^R_{\textrm{eq}}(\vec{r}, \vec{r}', \omega) - D^A_{\textrm{eq}}(\vec{r}, \vec{r}', \omega)\biggr] \text{,}
\end{split}
\end{equation}
and it has both real and imaginary parts, which are related by a Hilbert transform,
\begin{subequations}
\begin{align}
    \Re[K(\vec{r}, \vec{r}', t)] &= -\partial_t^2 \frac{D^R_{\textrm{eq}}(\vec{r}, \vec{r}', t) - D^A_{\textrm{eq}}(\vec{r}, \vec{r}', t)}{2} \text{,} \\
    \Im[K(\vec{r}, \vec{r}', t)] &= -\frac{1}{\pi} \mathcal{P} \int_{-\infty}^{\infty} \d t_1 \, \frac{\Re[K(\vec{r}, \vec{r}', t_1)]}{t-t_1} \text{.}
\end{align}
\end{subequations}
The symbol $\mathcal{P}$ here denotes the Cauchy principal value. 

From the presence of $\vec{E}$ on the right-hand sides of Eqs.~\eqref{eq:pre_ls_plus} and \eqref{eq:pre_ls_minus}, it is observed that the positive-frequency or negative-frequency part of the operator mixes both creation and annihilation operators, as previously claimed. 
By adding Eqs.~\eqref{eq:pre_ls_plus} and \eqref{eq:pre_ls_minus}, the imaginary part of $K$ cancels and the real part of $K$ adds up. 
The step function in $D^A$ ensures that it does not contribute, as $t_1 < t$ holds across the integration range. 
Thus, 
\begin{equation}
\begin{split}
    &\vec{E}(\vec{r}, t) 
    = \vec{E}_{\textrm{eq}}(\vec{r}, t) + \int \d^3 \vec{r}_1 \int_{-\infty}^{t} \d t_1 \, \\
    &\times \partial_t^2 D^R_{\textrm{eq}}(\vec{r}, \vec{r}_1, t-t_1) \epsilon_0 \chi_{\textrm{dr}}(t_1) \vec{E}(\vec{r}_1, t_1) \text{.} 
\end{split}
\end{equation}
By restoring the upper limit of the integral and using integration by parts, the Lippmann-Schwinger equation, Eq.~\eqref{eq:ls_e}, is recovered. 
Both NEGF and MQED are seen to arrive at the same conclusion, showing that the two theories are rigorously compatible with each other. 
\section{Discussion} \label{sec:discussion}

\subsection{Positive-frequency part of operator}

The definition of the positive frequency component of an operator is straightforward in the equilibrium case. 
One takes the Fourier transform of the function into the frequency domain, and keeps only the portion where $\omega > 0$. 
In quantum field theory, the positive frequency part of a field operator coincides with its annihilation component, as determined by the structure of the mode expansion. 
When time translational invariance is broken, the usual notion of positive frequency becomes ambiguous, because the very idea of frequency relies on a well-defined Fourier decomposition in time. 
In the periodic case, some components originally from the positive-frequency side might even get Floquet-shifted into the negative-frequency side. 
Here, we explore various possibilities for extending the definition of the positive frequency projection to the time-modulated case.

Building on the quantum field idea, the first possibility is to split the operator according to whether it acts as a creation or annihilation operator. 
Thus, we write $\vec{E} = \vec{E}^{[+]} + \vec{E}^{[-]}$, such that $\vec{E}^{[+]}$ contains only $\vec{a}^\omega_{0}$, and $\vec{E}^{[-]}$ contains only $(\vec{a}^\omega_{0})^\dagger$. 
We refer to this decomposition as the creation-annihilation split, motivated by normal-ordering considerations to subtract quantum vacuum contributions. 
By a linearity argument, both parts should satisfy the same Lippmann-Schwinger equation, 
\begin{equation}
\begin{split}
    &\vec{E}^{[\pm]}(\vec{r}, t) 
    = \vec{E}^{[\pm]}_{\textrm{eq}}(\vec{r}, t) + \int \d^3 \vec{r}_1 \int_{-\infty}^{\infty} \d t_1 \, \\
    & \times D^R_{\textrm{eq}}(\vec{r}, \vec{r}_1, t-t_1)[\partial_{t_1}^2  \epsilon_0 \chi_{\textrm{dr}}(\vec{r}_1, t_1)] \vec{E}^{[\pm]}(\vec{r}_1, t_1) 
    \text{.} \label{eq:e_pm_bracket_ls}
\end{split} 
\end{equation}

The second possibility, $E = E^{\{+\}} + E^{\{-\}}$, genuinely preserves the notion of ``positive frequency'', by employing the Sokhotski-Plemelj formula for the Fourier transform of the step function, 
\begin{equation}
    E^{\{\pm\}}(t) = \frac{1}{2} E(t) \pm \frac{i}{2\pi} \mathcal{P} \int \d t' \, \frac{E(t')}{t'-t} \text{.} 
\end{equation}
The drawback is that the second term requires knowing the behavior of the operator for all time, which is usually infeasible for time-dependent problems.

The third possibility is to define the split $\vec{E} = \vec{E}^{(+)} + \vec{E}^{(-)}$ according to Eqs.~\eqref{eq:mqed_e_plus} and \eqref{eq:mqed_e_minus}. 
We refer to this decomposition as the positive-frequency part as defined by the Heisenberg equation for $\vec{a}^{\omega}(\vec{r}, t)$. 
In this intermediate approach, we have seen that the creation and annihilation operators become mixed.
Although all three definitions are undoubtedly distinct, they agree when there is no time modulation.

\subsection{Intensity as a quantum correlation}

Having established that there are various definitions for the positive-frequency part of an operator, the question arises as to which is the most appropriate. 
In the context of thermal emission spectrum, the quantities of interest are typically the EM energy density and the Poynting vector. 
Both depend on field correlators, and by invoking Maxwell’s equations, it suffices to evaluate the electric field correlation $\langle \vec{E} \vec{E} \rangle$. 
Unlike the equilibrium scenario, the power emitted in the driven case is not constant in time, and thus, we consider the average taken over one period of the modulation.
The time average in the Floquet representation is given by
\begin{equation}
    \frac{\Omega}{2\pi} \int_{0}^{\frac{2\pi}{\Omega}} \d t \, D(t,t) = \int_{-\frac{\Omega}{2}}^{\frac{\Omega}{2}} \frac{\d \omega}{2\pi} \, \sum_m D_{mm}(\omega) 
    \text{.} \label{eq:time_average_floquet}
\end{equation}
The integral limits from $-\Omega/2$ to $\Omega/2$ reflect the Floquet Brillouin zone inherent to the Floquet representation. 
Alternatively, one can take the 00 block and integrate over all real frequencies. 
Before specifying the precise meaning of this quantum expectation value $\langle \vec{E} \vec{E} \rangle$, we first examine the physical conditions that are expected.

An essential criterion for a frequency spectrum is that it be integrable. 
In their previous work~\cite{vazquez-lozano_incandescent_2023}, Vazquez-Lozano et al. were faced with a non-vanishing tail that behaves like $N+1$ as $\omega \to \infty$.
The problematic ``$1$'' term leading to divergence can be regarded as a manifestation of the quantum vacuum and must be subtracted to yield physically meaningful results. 
This claim is further corroborated by comparing with Planck's radiation law for blackbodies in equilibrium, where the power spectrum is proportional to $N$, rather than $N+1$. 
In quantum field theory, vacuum contributions are systematically removed through the procedure of normal ordering.
Separating operators into annihilation and creation components naturally lends itself to the normal ordering procedure. 
Continuing with the previous notation that the superscript $[+]$ indicates the positive-frequency part associated with the annihilation operator, the normal-ordered product can be reformulated using the positive- and negative-frequency parts as 
\begin{equation}
    \langle : \vec{E} \vec{E} : \rangle = 2 \Re \langle \vec{E}^{[-]} \vec{E}^{[+]} \rangle \text{.} \label{eq:normal_ordered_product}
\end{equation}

Although Eq.~\eqref{eq:normal_ordered_product} appears to be a suitable candidate for the expression of the physically observed electric field intensity, a somewhat unsatisfactory point is revealed upon further analysis. 
At higher temperatures, thermal fluctuations may mask delicate quantum statistics and time-modulated effects. 
This leads us to a fundamental question: Does thermal emission persist in systems subject to temporal modulation even when the reservoir temperature is zero?
Several authors~\cite{vazquez-lozano_incandescent_2023,yu_near-field_2025} have noted that quantum fluctuations can surpass thermal fluctuations even at room temperature.
The energy injected by the active temporal modulation necessitates a corresponding channel for redistribution, excitation, or emission.
As such, we expect $\langle \vec{E} \vec{E} \rangle$ to not vanish even in the limit $T \to 0$. 
Unfortunately, $\langle \vec{E}^{[-]} \vec{E}^{[+]} \rangle$ vanishes as $T \to 0$, which can be seen by linking it to $\langle \vec{E}^{[-]}_{\textrm{eq}} \vec{E}^{[+]}_{\textrm{eq}} \rangle$ via the Lippmann-Schwinger equation, Eq.~\eqref{eq:e_pm_bracket_ls}.

Taking into account the preceding considerations, we propose to define $\langle \vec{E} \vec{E} \rangle$ as the symmetrized product integrated over the positive frequencies.
In the NEGF language, this corresponds to using the symmetrized GF $D^K = D^< + D^>$. 
Since $D^K$ is symmetric in $\omega$, the integration only needs to be carried out over positive frequencies to calculate the time average. 
The use of $D^K$ produces an $(N + 1/2)$ factor consistent with the fluctuation--dissipation theorem in FE~\cite{yu_manipulating_2023}.
However, the $1/2$ factor causes the aforementioned divergence issue, and should therefore be omitted.
The resulting expression is therefore proportional to $N$, as required for consistency with Planck’s radiation law.
From an implementation point of view, this is equivalent to working with the real part of $i\hbar D^<$ over the positive frequencies. 
As a sanity check, we confirm that this procedure reproduces the correct result in the static equilibrium case. 
To better understand why using $D^<$ give rise to the dynamic vacuum effect at $T \to 0$, we proceed by exploring the perturbative expansion of $D^<$.

\subsection{Formulae by perturbative orders}

To elucidate the structure in the perturbation expansion, we organize the terms following the notation in Ref.~\cite{vazquez-lozano_incandescent_2023}. 
In view of Eq.~\eqref{eq:ee_from_dless}, our quantity of interest is
\begin{equation}
    \mathcal{S}(\omega) = 2 \Re i \hbar \Tr \bm{\Omega}^2 \bm{D}^{<} 
    \text{.} \label{eq:define_s}
\end{equation}
According to Eq.~\eqref{eq:keldysh_truncated}, we have the following decomposition of $\mathcal{S}$,
\begin{equation}
    \mathcal{S} = \underbrace{\mathcal{S}_{0,0}}_{\mathcal{S}_0} 
    + \underbrace{\mathcal{S}_{1,0} + \mathcal{S}_{0,1}}_{\mathcal{S}_1} 
    + \underbrace{\mathcal{S}_{2,0} + \mathcal{S}_{1,1} + \mathcal{S}_{0,2}}_{\mathcal{S}_2} 
    + \ldots 
    \text{,}
\end{equation}
where the subscript $(a,b)$ indicates that there are $a$ and $b$ copies of $\Pi_{\textrm{dr}}$ on the right and left of $D^{<}_{\textrm{eq}}$ respectively. 

To the zeroth order, we have
\begin{equation}
\begin{split}
    &\mathcal{S}_{0,0}(\vec{r}, \omega) = -2 \hbar \Tr \Im  \bm{\Omega}^2 D^{<}_{\textrm{eq}} \\
    &= 4 \epsilon_0 \hbar \omega^4 N(\omega) \Im[\epsilon_{\textrm{eq}}(\omega)] \Re \Tr D^R_{\textrm{eq}}(\omega) D^A_{\textrm{eq}}(\omega) 
    \text{.} \label{eq:s00_simplified}
\end{split}
\end{equation} 
The spatial convolutions in Eq.~\eqref{eq:s00_simplified} are present but omitted from display for clarity, just as in Eq.~\eqref{eq:dyson_ls}. 
When written out in full, the convolution reads
\begin{equation}
    \Tr D^R_{\textrm{eq}} D^A_{\textrm{eq}} \equiv \int_{\rho_z < 0} \d^{3} \vec{\rho} \, \Tr[ D^R_{\textrm{eq}}(\vec{r}, \vec{\rho}, \omega) D^A_{\textrm{eq}}(\vec{\rho}, \vec{r}, \vec{\omega}) ] \text{.}
\end{equation}
The effect of time modulation is absent in the zeroth-order term, and the fluctuation--dissipation result is recovered.

Next, we briefly reason that the first-order terms do not contribute. 
They are defined by
\begin{subequations}
\begin{align}
    \mathcal{S}_{0,1} &= -2 \hbar \Im \Tr \bm{\Omega}^2 \bm{D}^R_{\textrm{eq}} \bm{\Pi}_{\textrm{dr}} \bm{D}^{<}_{\textrm{eq}} = 0 \text{,} \label{eq:s01} \\
    \mathcal{S}_{1,0} &= -2 \hbar \Im \Tr \bm{\Omega}^2 \bm{D}^{<}_{\textrm{eq}} \bm{\Pi}_{\textrm{dr}} \bm{D}^A_{\textrm{eq}} = 0 \text{.} \label{eq:s10} 
\end{align}
\end{subequations}
All Floquet matrices in Eqs.~\eqref{eq:s01} and \eqref{eq:s10} are diagonal, except for $\bm{\Pi}_{\textrm{dr}}$ which has zero along the main diagonal.
Multiplication by a diagonal matrix amounts to scaling of the columns/rows. 
Therefore, the final product will also be zero along the main diagonal, and in particular, at the 00 block.
More generally, similar reasoning implies that odd-order terms do not contribute at higher orders in perturbation theory. 

Consequently, the leading-order perturbative effects arise at second order, which consists of three terms given by
\begin{subequations}
\begin{align}
    \mathcal{S}_{0,2} &= -2 \hbar \Im \Tr \bm{\Omega}^2 \bm{D}^R_{\textrm{eq}} \bm{\Pi}_{\textrm{dr}} \bm{D}^R_{\textrm{eq}} \bm{\Pi}_{\textrm{dr}} \bm{D}^{<}_{\textrm{eq}} \text{,} \label{eq:s02} \\
    \mathcal{S}_{2,0} &= -2 \hbar \Im \Tr \bm{\Omega}^2 \bm{D}^{<}_{\textrm{eq}} \bm{\Pi}_{\textrm{dr}} \bm{D}^A_{\textrm{eq}} \bm{\Pi}_{\textrm{dr}} \bm{D}^A_{\textrm{eq}} \text{,} \label{eq:s20} \\
    \mathcal{S}_{1,1} &= -2 \hbar \Im \Tr \bm{\Omega}^2 \bm{D}^R_{\textrm{eq}} \bm{\Pi}_{\textrm{dr}} \bm{D}^{<}_{\textrm{eq}} \bm{\Pi}_{\textrm{dr}} \bm{D}^A_{\textrm{eq}} \text{.} \label{eq:s11} 
\end{align}
\end{subequations}
Upon expanding using Eq.~\eqref{eq:pi_dr_floquet} and simplifying, we have  
\begin{subequations}
\begin{align}
\begin{split}
    \mathcal{S}_{0,2} &= 4 \Delta \chi^2 \epsilon_0^3 \hbar \omega^6 N(\omega) \Im[\epsilon_{\textrm{eq}}(\omega)] \sum_{n=\pm 1} \omega_n^2 \\
    &\quad \times \Re \Tr D^R_{\textrm{eq}}(\omega) D^R_{\textrm{eq}}(\omega_n) D^R_{\textrm{eq}}(\omega) D^A_{\textrm{eq}}(\omega) \text{,} \label{eq:s02_simplified}
\end{split}
\\
\begin{split}
    \mathcal{S}_{2,0} &= 4 \Delta \chi^2 \epsilon_0^3 \hbar \omega^6 N(\omega) \Im[\epsilon_{\textrm{eq}}(\omega)] \sum_{n=\pm 1} \omega_n^2 \\
    &\quad \times \Re \Tr D^R_{\textrm{eq}}(\omega) D^A_{\textrm{eq}}(\omega) D^A_{\textrm{eq}}(\omega_n) D^A_{\textrm{eq}}(\omega) \text{,} \label{eq:s20_simplified}
\end{split} 
\\
\begin{split}
    \mathcal{S}_{1,1} &= 4 \Delta \chi^2 \epsilon_0^3 \hbar \omega^4 \sum_{n=\pm 1} N(\omega_n) \Im[\epsilon_{\textrm{eq}}(\omega_n)] \omega_n^4 \\
    &\quad \times \Re \Tr D^R_{\textrm{eq}}(\omega) D^R_{\textrm{eq}}(\omega_n) D^A_{\textrm{eq}}(\omega_n) D^A_{\textrm{eq}}(\omega) \text{.} \label{eq:s11_simplified}
\end{split}
\end{align}
\end{subequations}

Since $\mathcal{S}_{0,2} = \mathcal{S}_{2,0}$, the evaluation of either term alone is adequate, and we write $\mathcal{S}_2 = 2 \mathcal{S}_{0,2} + \mathcal{S}_{1,1}$. 
It is worth pointing out how the dynamical vacuum effects manifest in Eq.~\eqref{eq:s11_simplified}. 
For the case of $n = -1$, the Bose-Einstein factor is right-shifted by $\Omega$, such that in the $T \to 0$ limit, we have $N(\omega) \to -1$ for $-\Omega < \omega < 0$. 
This nonvanishing factor renders $\mathcal{S}_{1,1}$ the leading-order term responsible for the quantum vacuum effect. 
To facilitate comparison, Table~\ref{tab:formula_comparison} below lists a one-to-one correspondence between our results and those given in the Supplementary Information of Ref.~\cite{vazquez-lozano_incandescent_2023}. 
\begin{table}[htbp]
  \caption{Comparison of perturbation expansion formulae in this work with those in Ref.~\cite{vazquez-lozano_incandescent_2023}}
  \label{tab:formula_comparison}
  \centering
  \setlength{\tabcolsep}{10pt} % wider columns
  \begin{tabular}{ccc}
    \hline
    Quantity & This work & Ref.~\cite{vazquez-lozano_incandescent_2023} \\
    \hline
    $S_{0,0}$ & Eq.~\eqref{eq:s00_simplified} & Eq.~(S.VI.14) \\
    $S_{0,2}$ & Eq.~\eqref{eq:s02_simplified} & Eq.~(S.VI.29) \\
    $S_{1,1}$ & Eq.~\eqref{eq:s11_simplified} & Eq.~(S.VI.32) \\
    \hline
  \end{tabular}
\end{table}
\setlength{\tabcolsep}{6pt}  % reset to APS default

\subsection{Numerical results}

\begin{figure}[htb]
    \centering
    \includegraphics[width=\linewidth]{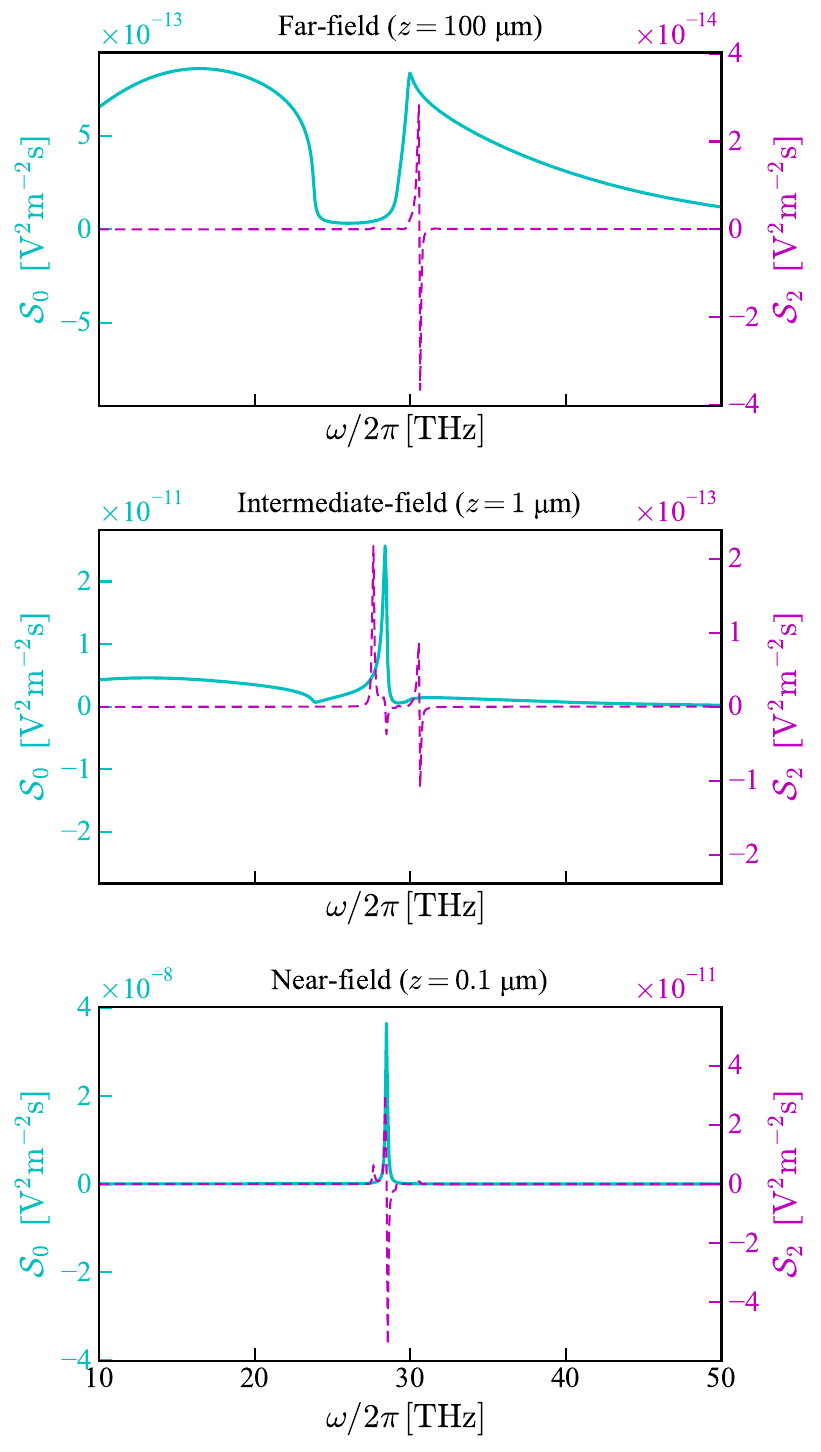}
    \caption{
    \label{fig:numerical} 
    Frequency spectra of the electric field correlation measured at three distances from the slab. 
    The blue solid curves correspond to the equilibrium value $\mathcal{S}_0$, while the purple dashed curves represent $\mathcal{S}_2$, the leading order correction due to the time modulation. 
    To facilitate comparison, we adopt the same parameters as those used in Fig. 5 of Ref.~\cite{vazquez-lozano_incandescent_2023}. 
    }
\end{figure}

From an implementation standpoint, the principal difficulty lies in evaluating successive convolutions of the GFs.
Fortunately, the technical groundwork has been laid out in Ref.~\cite{vazquez-lozano_incandescent_2023}, and we will directly adopt their results, with only minor modifications to correct for typographical errors. 
In Fig.~\ref{fig:numerical}, the zeroth-order and second-order contributions of $\langle \vec{E} \cdot \vec{E} \rangle$ are plotted using dual vertical axes, enabling visible juxtaposition despite differing magnitudes. 
Most of the salient features are captured in the range of 10--50~\si{\tera\hertz},  with the exponential decay of the tail particularly evident in the far-field regime.
We observe that the secondary contribution from $\mathcal{S}_2$ is suppressed by 1--3 orders of magnitude relative to $\mathcal{S}_0$, in marked contrast to Fig.~5 of Ref.~\cite{vazquez-lozano_incandescent_2023} where $\mathcal{S}_2$ was shown to play a more prominent role.
Consequently, the inclusion of $\mathcal{S}_2$ in the total spectrum $\mathcal{S}$ does not lead to a discernible enhancement, let alone surpass the blackbody spectrum.  

Another important feature of the spectra is the exponential decay of the tails in the high-frequency limit. 
As noted earlier, this behavior originates from the Bose-Einstein factor present in $D^<_{\textrm{eq}}$ due to the fluctuation--dissipation theorem. 
In Ref.~\cite{vazquez-lozano_incandescent_2023}, a major conceptual difficulty arises when integrating the spectrum over all frequencies, which is mathematically divergent due to the non-vanishing tail in their $\mathcal{S}_{1,1}$ term. 
To address this spectral pathology, the authors in a follow-up work~\cite{vertiz-conde_dispersion_2025} proposed a memory-time parameter to effectuate a cut-off at high frequencies. 
Here, we demonstrate that such additional assumptions are unnecessary. 
From our earlier analysis, this divergence can be eliminated if we choose to use $D^<$ instead of $\langle E^{(-)} E^{(+)} \rangle$ (positive-frequency part defined by the Heisenberg equation). 
It is due to the mixing of $\vec{a}^{\omega}_{0}$ and $(\vec{a}^{\omega}_{0})^\dagger$ that results in terms with asymptotic behaviour of $(N+1)$ that do not decay. 
Fortunately, the majority of the results in Ref.~\cite{vazquez-lozano_incandescent_2023} remain applicable.
Borrowing their notation, the only effective changes to be made is to discard the $\tilde{\mathcal{G}}_{1,1}^{(\omega^\pm)}$ term in $\mathcal{S}_{1,1}^{(\omega^\pm)}$ while retaining the $\bar{\mathcal{G}}_{1,1}^{(\omega^\pm)}$ term. 
This $\tilde{\mathcal{G}}_{1,1}^{(\omega^\pm)}$ term is primarily responsible for the ``vacuum amplificaton effect'' discussed there. 
The dramatic decrease in the enhancement effect can be understood by computing the ratio of $N/(N+1)$ near the resonance frequency of $\SI{29}{\tera\hertz}$, which turns out to be only about 1\% at room temperature. 
Thus, by excluding the problematic $(N+1)$ term, the divergence issue is resolved, albeit at the expense of the dominant contribution to the second-order correction.

From a broader perspective, we should carefully contemplate which version of $\langle \vec{E} \cdot \vec{E} \rangle$ agrees with experimental measurements, and it seems reasonable to suggest that a well-considered definition should not diverge. 
Although certain distinctions can be attributed to definitions, others reflect fundamental differences that cannot be reduced to a choice of convention.
For instance, our formulas agree almost perfectly with Ref.~\cite{vazquez-lozano_incandescent_2023} up to a factor of $\pi$, which is purely a normalization convention with no physical significance. 
Yet, based on the Lippmann-Schwinger equation, we suggest that the interaction term in the MQED Hamiltonian arising from the time modulation should be $-\tfrac{1}{2} \vec{E} \cdot \vec{P}_{\textrm{dr}}$ instead of $-\vec{E} \cdot \vec{P}_{\textrm{dr}}$. 
The factor of $1/2$ translates into a $1/4$ reduction in the already weak $\mathcal{S}_2$, thereby exacerbating the challenge of its experimental detection. 
\section{Conclusion}

In summary, we applied the NEGF formalism to investigate the impact of time modulation in SiC on its intensity spectrum. 
In parallel, we demonstrated how the MQED Lagrangian can be adapted to incorporate such modulation to model lossless variations in the permittivity.
We rigorously established the compatibility of both approaches by showing that they independently yield the same Lippmann-Schwinger equation.
We propose multiple approaches for decomposing the electric field into positive- and negative-frequency components, which motivates the use of the positive-frequency part of the lesser photon GF to calculate $\langle \vec{E} \vec{E} \rangle$ that is consistent with physical expectations.
From this definition, we show--both analytically and numerically--that there is no divergence issue, although the resulting enhancement effect is minimal. 
Taken together, our results lay the groundwork for more targeted investigations into time-modulated media and underscore the continued relevance of Floquet engineering as a framework for uncovering novel physical phenomena.

\section{Acknowledgements}

We acknowledge support from the Ministry of Education, Singapore, under the Academic Research Fund (FY2022).

\bibliographystyle{apsrev4-2}

\end{document}